\documentclass[aps,superscriptaddress,showpacs,twocolumn]{revtex4}
\usepackage[dvips]{graphicx}
\usepackage{amssymb}

\begin{document}
\title{Thermodynamic phase transitions for Pomeau-Manneville maps}

\author{Roberto Venegeroles}\email{roberto.venegeroles@ufabc.edu.br}
\address{Centro de Matem\'atica, Computa\c c\~ao e Cogni\c c\~ao, UFABC, 09210-170, Santo Andr\'e, SP, Brazil}

\date{\today}

\begin{abstract}
We study phase transitions in the thermodynamic description of Pomeau-Manneville intermittent maps from the point of view of infinite ergodic theory, which deals with diverging measure dynamical systems. For such systems, we use a distributional limit theorem to provide both a powerful tool for calculating thermodynamic potentials as also an understanding of the dynamic characteristics at each instability phase. In particular, topological pressure and R\'enyi entropy are calculated exactly for such systems. Finally, we show the connection of the distributional limit theorem with non-Gaussian fluctuations of the algorithmic complexity proposed by Gaspard and Wang [Proc. Natl. Acad. Sci. U.S.A. {\bf 85}, 4591 (1988)].
\end{abstract}


\maketitle

\section{Introduction}

The major goal of statistical mechanics is to explain the macroscopic properties of complex systems in terms of a very small number of parameters by using probabilistic approaches. As is well known, its primary motivation was the study of thermodynamical properties of matter based on random behavior of their very large number of constituents (atom and molecules). Almost a century after Boltzmann's seminal work \cite{Boltz}, such approaches were extended to dynamical systems theory \cite{Sinai,Bowen,Ruelle}, a branch that is currently known as thermodynamic formalism \cite{Ruelle}. In this scenario, the chaotic dynamics of an ensemble of trajectories plays the role of randomness in the many-body dynamics, even for one degree-of-freedom dynamical systems. The thermodynamic approach has proven to be a powerful tool in the ergodic theory of hyperbolic and expanding dynamical systems \cite{Ruelle}. Later, there has been growing interest, mostly by theoretical physicists, in extending this approach to more general dynamical systems (see \cite{BS} and references. therein), particularly those that exhibit fractal sets \cite{FJP,KP,BJ,BP,GBP} or some kind of intermittent behavior \cite{STCK,Wang,Feign,MKHH,SH,Prell}.

Here we deal with phase transitions for Pomeau-Manneville (PM) maps $x_{t+1}=f(x_{t})$ where  $f$ takes the form
\begin{eqnarray}
\label{PMmap}
f(x)=x(1+ax^{1/\alpha})\,\,\,\mbox{mod}\,1,
\end{eqnarray}
with $a > 0$ and $\alpha > 0$ \cite{PM}. The remarkable characteristic of such systems is the intermittent behavior due to the presence of the indifferent fixed point $x=0$, i.e., $f(0)=0$ and $f'(0)=1$. It is important to stress that the global form of $f$ far from $x=0$ is less relevant here. For example, systems behaving like (\ref{PMmap}) on $[0,x_{*})$, where $f(x_{*})=1$, exhibit the same statistical behavior of (\ref{PMmap}) since the map on $[x_{*},1]$ is given by some well-behaved function $f_{1}$ such that $f_{1}(x_{*})=0$ and $f_{1}(1)=1$. Systems of the type (\ref{PMmap}) have diverging invariant measure $\mu(x)$ near their indifferent fixed points for $0<\alpha<1$. More specifically, the invariant density $\omega(x)$ of map (\ref{PMmap}), where  $d\mu(x)=\omega(x)dx$, behaves as
\begin{eqnarray}
\label{omeg}
\omega(x)\sim bx^{-1/\alpha},
\end{eqnarray}
near $x=0$ \cite{Thaler}. Therefore, diverging measure regime of (\ref{omeg}) leads to a very slow laminar phase near $x=0$ alternating with fast turbulent one elsewhere. Due to this peculiarity, the dynamics of the system (\ref{PMmap}) exhibit subexponential instability of the type $|\delta x_{t}|\sim|\delta x_{0}| \exp(\lambda_{\alpha} t^{\alpha})$ for $0<\alpha<1$ \cite{SV}. On the other hand, $\alpha>1$ leads to the finiteness of invariant measure, which is naturally related to the usual chaos and ordinary Lyapunov exponents.

It is important to point out here the connection between subexponential instability and the so-called
``sporadic randomness,'' a phenomenon initially studied by Gaspard and Wang \cite{GW}. These authors conjectured that the Kolmogorov-Chaitin algorithmic complexity $C_{t}$ for map (\ref{PMmap}) is proportional to the number of entrances $N_{t}$ into a given phase space cell after a large number of iterations. In this assumption, recently confirmed in \cite{SV} by means of a Pesin-type indentity, the statistics of $N_{t}$ is ruled by non-Gaussian fluctuations involving Feller's renewal results \cite{Feller}. Subsequently, thermodynamic phase transitions of PM map (\ref{PMmap}) for $0<\alpha<2$ was studied by Wang \cite{Wang} employing the same approach of \cite{GW}. It is also interesting to note that the sporadic randomness has not only been verified in PM intermittent maps (e.g., \cite{SV}), but also suggested as a distinguishing feature in weather systems \cite{RPNEB}, noncoding DNA sequences \cite{LK}, and some linguistic texts \cite{EN}.

The purpose of this work is twofold. First, we will revisit the pioneering results of \cite{Wang}, but now from the point of view of infinite ergodic theory \cite{Aaronson} (for some applications, see also \cite{SV,PSV}). In this first part some results involving phase transitions of the so-called topological pressure are considerably improved. The topological pressure can be interpreted as a free energy density associated with the ensemble of trajectories. We also discuss the phase transition related to the R\'enyi entropy, extending the results observed in \cite{STCK} to the diverging measure (nonergodic) regime of PM map (\ref{PMmap}). Finally, we show the connection between Feller's sporadic statistics and the infinite ergodic theory. Second, it aims at understanding the phase transition problem from a dynamical point of view since singular behavior of thermodynamical quantities does not tell everything about dynamic characteristics of a system. The approach employed here also show us precisely what happens at each phase, particularly in the subexponential regime of map (\ref{PMmap}).

\section{Topological Pressure}

In the thermodynamic formalism, systems of the type (\ref{PMmap}) exhibit continuous phase transition, a situation where thermodynamic quantities vary continuously but not analytically when some external parameter of the system is changed. The paradigmatic example in the usual statistical mechanics is the ferromagnetic material at zero external magnetic field: Ferromagnets lose their spontaneous magnetization when heated above a specific critical temperature $T_{c}$ and the derivative of the magnetization with respect to magnetic field (susceptibility) diverges at $T_{c}$ and zero field. As shown in \cite{Wang}, the many-body model that most closely resembles (\ref{PMmap}) is the Fisher-Felderhof droplet model of condensation \cite{FF}.

Let us first consider the topological pressure $P(\beta)$, a kind of negative Helmholtz free energy of thermodynamic formalism, defined as \cite{BS}
\begin{eqnarray}
\label{Prestop}
P(\beta)=\lim_{t\rightarrow\infty}\frac{1}{t}\ln Z_{t}(\beta),
\end{eqnarray}
with the corresponding partition function $Z_{t}(\beta)$ given by
\begin{eqnarray}
\label{partic}
Z_{t}(\beta)=\sum_{\left\{x_{i}\right\}}\exp\left[-\beta\sum_{k=0}^{t-1}\ln|f'(f^{k}(x_{i}))|\right].
\end{eqnarray}
The set of points $\left\{x_{i}\right\}$ in Eq. (\ref{partic}) is chosen as follows. First, consider a partition of phase space into disjoint boxes $\Delta_{i}$ so that transitions between nearest-neighbor configurations $(i,i')$ are possible, i.e., $f(\Delta_{i})\cap\Delta_{i'}\neq\emptyset$. For each allowed sequence $i_{0},\ldots,i_{t-1}$, there is a subset $\Delta_{x}(i_{0},\ldots,i_{t-1})$ of phase space defined by
\begin{eqnarray}
\label{Delta}
\Delta_{x}(i_{0},\ldots,i_{t-1})=\left\{x:f^{k}(x)\in\Delta_{i_{k}},k=0,\ldots,t-1\right\}.\nonumber\\
\end{eqnarray}
The size of subsets $\Delta_{i_{k}}$ goes to zero as $t\rightarrow\infty$. Then, for very large but still finite $t$, we pick a representative point $x_{i}$, one from each subset, and collect them as the set $\left\{x_{i}\right\}$. It is important to stress, however, that the analytical determination of this set is not usually a practical task. We can circumvent this problem by replacing the summation over $x_{i}$ by the integration over the conditional measure as follows ($h$ is an arbitrary function)
\begin{eqnarray}
\label{sumint}
\sum_{\left\{x_{i}\right\}}h(x_{i})\sim\int_{x\in[0,1]}d\sigma(x)|\Delta_{x}(i_{0},\ldots,i_{t-1})|h(x),
\end{eqnarray}
where $\sigma(x)$ represents the measure of initial condition $x$ and $|\Delta|$ denotes the Lebesgue measure of $\Delta$. As $t\rightarrow\infty$ we have the following property \cite{EcP}
\begin{eqnarray}
\label{Leb}
|\Delta_{x}(i_{0},\ldots,i_{t-1})|=|\Delta_{x}(i_{1},\ldots,i_{t-1})||f'(x)|,
\end{eqnarray}
where, on the right side of (\ref{Leb}), we can replace $i_{1}$ by $i_{2}$ and $|f'(x)|$ by $|f'[f(x)]||f'(x)|$, and so forth, leading to
\begin{eqnarray}
\label{Lebdel}
|\Delta_{x}(i_{0},\ldots,i_{t-1})|=\prod_{k=0}^{t-1}|f'[f^{k}(x)]|.
\end{eqnarray}
Finally, we can rewrite Eq. (\ref{partic}) by means of Eqs. (\ref{sumint}) and (\ref{Lebdel}) yielding
\begin{eqnarray}
\label{Zint}
Z_{t}(\beta)\sim\int d\sigma(x)\exp\left[(1-\beta)\sum_{k=0}^{t-1}\ln|f'[f^{k}(x)]|\right].
\end{eqnarray}

Before attempting to estimate Eq. (\ref{Zint}) we will make use of the Aaronson-Darling-Kac (ADK) theorem \cite{Aaronson}, which is precisely applicable to PM systems of type (\ref{PMmap}). For such systems, this theorem ensures that, for a positive function $\vartheta$ integrable over $\mu$ and an arbitrary measure $\sigma$ of initial conditions absolutely continuous with respect to the Lebesgue measure, we have
\begin{eqnarray}
\label{ADK}
\frac{1}{t^{\gamma}}\sum_{k=0}^{t-1}\vartheta[f^{k}(x)]\stackrel{d}{\rightarrow}\xi_{\gamma}c_{\gamma}(t)\int\vartheta d\mu,
\end{eqnarray}
as $t\rightarrow\infty$, where $\xi_{\gamma}$ is a non-negative Mittag-Leffler random variable of index $\gamma\in(0,1]$ and expected value $E(\xi_{\gamma})=1$. The corresponding Mittag-Leffler probability density function $\rho_{\gamma}(\xi)$ is given by \cite{Feller,SVml}
\begin{eqnarray}
\label{ML}
\rho_{\gamma}(\xi)=\frac{\Gamma^{1/\gamma}(1+\gamma)}{\gamma\xi^{1+1/\gamma}}g_{\gamma}\left[\frac{\Gamma^{1/\gamma}(1+\gamma)}{\xi^{1/\gamma}}\right],
\end{eqnarray}
where $g_{\gamma}$ stands for the one-sided L\'evy stable density, whose Laplace transform is $\tilde{g}(u)=\exp(-u^{\gamma})$ (see \cite{SVml,PG} for a detailed discussion). For PM maps of the type (\ref{PMmap}), the index $\gamma$ is
\begin{eqnarray}
\label{gam}
\gamma = \left\{
\begin{array}{ll}
\displaystyle \alpha, & 0<\alpha<\displaystyle1,\\
\displaystyle 1, & \displaystyle\alpha\geq1,
\end{array}
\right.
\end{eqnarray}
whereas the coefficient $c_{\gamma}(t)$ in Eq. (\ref{ADK}) takes the asymptotic form \cite{RZ}
\begin{eqnarray}
\label{ct}
c_{\gamma}(t)\sim\left\{
\begin{array}{ll}
\displaystyle \frac{1}{ba}\left(\frac{a}{\alpha}\right)^{\alpha}\frac{\sin(\pi\alpha)}{\pi\alpha}, & 0<\alpha<\displaystyle1,\\
\displaystyle (b\ln t)^{-1}, & \displaystyle\alpha=1,\\
\displaystyle 1, & \displaystyle\alpha>1,
\end{array}
\right.
\end{eqnarray}
as $t\rightarrow\infty$, recalling that $b=\lim_{x\rightarrow0}x^{1/\alpha}\omega(x)$. For $\alpha>1$ we have introduced the Birkhoff ergodic case $\gamma=1$, for which the corresponding Mittag-Leffler density reduces to $\rho_{1}(\xi)=\delta(1-\xi)$, as in the $\alpha=1$ case. Evidently, we can choose $\vartheta=\ln|f'|$ in the ADK formula (\ref{ADK}).

Consider now the algorithmic complexity $C_{t}$ of PM map (\ref{PMmap}), valid for all $\alpha>0$ \cite{SV}:
\begin{eqnarray}
\label{Comp}
C_{t}(x)\sim\sum_{k=0}^{t-1}\ln|f'[f^{k}(x)]|,
\end{eqnarray}
as $t\rightarrow\infty$. Equations (\ref{ADK}) and (\ref{ct}) lead to (see also \cite{SV})
\begin{eqnarray}
\label{CompML}
\frac{C_{t}}{\left\langle C_{t}\right\rangle}\stackrel{d}{\rightarrow}\xi_{\gamma},
\end{eqnarray}
as $t\rightarrow\infty$, where the ADK average $\left\langle C_{t}\right\rangle=h_{\gamma}t^{\gamma}$ is given in terms of the average of generalized Kolmogorov-Sinai entropy \cite{SV}
\begin{eqnarray}
\label{hgam}
h_{\gamma}=c_{\gamma}\int d\mu\ln|f'|.
\end{eqnarray}
Going back to Eq. (\ref{Zint}), we can overcome the integration problem over arbitrary $\sigma$ considering it absolutely continuous with respect to the Lebesgue measure. Such condition is sufficiently broad to assure that our results involving phase transitions typically do not depend on the initial condition distributions. This is somewhat surprising in the case of nonergodic regimes, i.e., $0<\alpha<1$ in the present case. After applying Eqs. (\ref{CompML}) and (\ref{hgam}) in the ADK formula (\ref{ADK}), we have
\begin{eqnarray}
\label{zml}
Z_{t}(\beta)\sim\int_{0}^{\infty}d\xi\rho_{\gamma}(\xi)\exp[-(\beta-1)h_{\gamma}t^{\gamma}\xi].
\end{eqnarray}
Note that Eq. (\ref{zml}) is just the Laplace transform of $\rho_{\gamma}$, which is given by the Mittag-Leffler special function $E_{\gamma}(u)$, namely \cite{SVml}
\begin{eqnarray}
\label{MLsf}
\tilde{\rho_{\gamma}}(u)=E_{\gamma}(u)=\sum_{n=0}^{\infty}\frac{[\Gamma(1+\gamma)u]^{n}}{\Gamma(1+n\gamma)},
\end{eqnarray}
with $u=(1-\beta)h_{\gamma}t^{\gamma}$. Now, considering the asymptotes
 $E_{\gamma}(u)\sim\gamma^{-1}\exp(u^{1/\gamma})$ as $u\rightarrow\infty$ \cite{ML} and $E_{\gamma}(u)\sim0$ as $u\rightarrow-\infty$ \cite{note}, we have finally for all $\alpha>0$ and $\beta$ near $1$
\begin{eqnarray}
\label{PTph}
P(\beta) \sim \left\{
\begin{array}{ll}
\displaystyle [h_{\gamma}(1-\beta)]^{1/\gamma}, & \displaystyle \beta<1,\\
\displaystyle 0, & \displaystyle\beta\geq1,
\end{array}
\right.
\end{eqnarray}
observing that $h_{\gamma}=0$ ($c_{\gamma}\rightarrow0$) for $\alpha=1$. Note that Eq. (\ref{PTph}) is in accordance with the results first obtained in \cite{Wang} for $0<\alpha<2$ and later extended for all $\alpha>0$ in \cite{Prell}. It is noteworthy here that, unlike these approaches, the prefactor $h_{\gamma}$ in Eq. (\ref{PTph}) is obtained exactly, given by Eq. (\ref{hgam}) for all $\alpha>0$.

\section{R\'enyi Entropy}

Let us consider now the phase transition related to the R\'enyi entropy \cite{BS}:
\begin{eqnarray}
\label{Renk}
K(\beta)=\frac{1}{1-\beta}\lim_{t \rightarrow\infty}\frac{1}{t}\ln\sum_{j=0}^{t-1}\left(p_{j}^{(t)}\right)^{\beta},
\end{eqnarray}
where $p_{j}^{(t)}=p_{j}(i_{0}, \ldots, i_{t-1})$ usually denotes the probability that a randomly chosen initial condition ($\sigma$ distributed) on the phase space falls into $\Delta_{j}$ at time $t-1$. In view of the fact that we are also dealing with nonergodic regimes ($0<\alpha<1$), we set $p_{j}^{(t)}$ such that
\begin{eqnarray}
\label{pbet}
p_{j}^{(t)}(q)=\frac{\tau_{j}^{q}}{\sum_{k=0}^{t-1}\tau_{k}^{q}},
\end{eqnarray}
where $p_{j}^{(t)}(q=1)=p_{j}^{(t)}$. In Eq. (\ref{pbet}) we consider the amount of time $\tau_{j}$ spent in state $\Delta_{j}$ instead of its length $|\Delta_{j}|$, which is usually considered for ergodic systems (for which $\tau_{j}\propto|\Delta_{j}|$). Then the partition function $Z_{t}$ takes the asymptotic form
\begin{eqnarray}
\label{ztq}
Z_{t}(q)\sim\exp[tP(q)]\sim\sum_{k=0}^{t-1}\tau_{k}^{q}.
\end{eqnarray}
Recalling that $\sum_{j}\left[p_{j}^{(t)}(q)\right]^{\beta}\sim\exp[(1-\beta)K(\beta,q)t]$, we have for $q=1$
\begin{eqnarray}
\label{kpb}
K(\beta)=\frac{P(\beta)-\beta P(1)}{1-\beta},
\end{eqnarray}
also valid for ergodic systems \cite{BS}. From the topological pressure (\ref{PTph}) we then have
\begin{eqnarray}
\label{Reny}
K(\beta) \sim \left\{
\begin{array}{ll}
\displaystyle h_{\gamma}^{1/\gamma}(1-\beta)^{-1+1/\gamma}, & \displaystyle \beta<1,\\
\displaystyle 0, & \displaystyle\beta\geq1.
\end{array}
\right.
\end{eqnarray}
Note that for $\gamma=1$, i.e., ergodic regimes, $K(\beta)=h_{KS}$ for $\beta<1$, where $h_{1}=h_{KS}$ is the Kolmogorov-Sinai entropy, whereas $K(\beta)=0$ for $\beta\geq1$. Therefore, Eq. (\ref{Reny}) extends for nonergodic regimes ($0<\alpha<1$) the nonanalytic behavior of R\'enyi entropy at $\beta=1$ observed in \cite{STCK}.

\section{Algorithmic complexity satisfies the ADK theorem}

In \cite{Wang}, as well as in \cite{GW}, the algorithmic complexity $C_{t}$ of a piecewise version of the PM map was considered as the random number of entrances $N_{t}$ into a given phase space cell ($A_{0}$) during $t$ iterations of the map, i.e., $C_{t}\sim N_{t}$. The statistics $p_ {\alpha}$ of $N_{t}$ employed is well known from Feller's renewal theorems \cite{Feller}, and it was applied in the estimation of $P(\beta)$. The accordance with Eq. (\ref{PTph}) can be understood by observing that $p_{\alpha}$ is, in fact, a Mittag-Leffler probability density function. The statistics of $N_{t}$ for the case $0<\alpha<1$ is given by \cite{Feller}
\begin{eqnarray}
\label{pg1}
P_{\alpha}\left(N_{t}\geq c_{1}\frac{t^{\alpha}}{q^{\alpha}}\right)\sim G_{\alpha}(q),
\end{eqnarray}
as $t\rightarrow\infty$, where $P_{\alpha}$ and $G_{\alpha}$ stand for the cumulative distribution functions of $p_{\alpha}$ and $g_{\alpha}$, respectively. Applying the change of variable $q=r\xi^{-1/\alpha}$, with $r^{\alpha}=\alpha\Gamma(\alpha)$ \cite{SVml}, and after introducing the normalized random variable $\xi=N_{t}/\left\langle N_{t}\right\rangle$, we have
\begin{eqnarray}
\label{pg2}
p_{\alpha}(N_{t})dN_{t}\sim\rho_{\alpha}(\xi)d\xi,
\end{eqnarray}
as $t\rightarrow\infty$, where $\left\langle N_{t}\right\rangle=c_{1}t^{\alpha}/\alpha\Gamma(\alpha)$ and $\rho_{\alpha}$ is the Mittag-Leffler density (\ref{ML}). For the case $\alpha>1$, but different from $2$, we have \cite{Feller}
\begin{eqnarray}
\label{pg3}
P_{\alpha}\left(N_{t}\geq c_{2}t-c_{3}t^{1/\kappa}q\right)\sim G_{\kappa}(q),
\end{eqnarray}
where $\kappa=\alpha$ for $1<\alpha<2$ and $\kappa=2$ for $\alpha>2$. Now $G_{\kappa}$ stands for the cumulative distribution function of the two-sided stable density $g_{\kappa}$ \cite{Wang}. We can consider the same normalized variable $\xi$ of Eq. (\ref{pg2}), but now with $\left\langle N_{t}\right\rangle=c_{2}t$. Then Eq. (\ref{pg3}) becomes
\begin{eqnarray}
\label{pg4}
p_{\alpha}(N_{t})dN_{t}\sim\frac{1}{\epsilon}g_{\kappa}\left(\frac{1-\xi}{\epsilon}\right)d\xi,
\end{eqnarray}
where $\epsilon=(c_{3}/c_{2})t^{-1+1/\kappa}$ goes to $0$ as $t\rightarrow\infty$. This leads to $\delta(1-\xi)d\xi$ on the right hand side of Eq. (\ref{pg4}). Already in the Gaussian case $\kappa= 2$, $\epsilon^{2}$ is proportional to the variance, also leading Eq. (\ref{pg4}) to the same Dirac $\delta$ function. The same occurs for $\alpha= 2$, where we also have Eq. (\ref{pg4}) for $\kappa=2$, but replacing $c_{3}$ by $c_{3}\sqrt{\ln t}$ and also leading to $\epsilon\rightarrow0$ as $t\rightarrow\infty$.

\section{FINAL REMARKS}

We revisit here the problem of thermodynamic phase transitions for PM maps (\ref{PMmap}) by using the infinite ergodic theory, in particular the ADK theorem. The topological pressure $P(\beta)$ and R\'enyi entropy $K(\beta)$ are calculated exactly for such systems, exhibiting both phase transitions at the same critical value $\beta_{c}=1$. Our results also shed some light on the role of the measure of initial conditions $\sigma$ in the calculation of these thermodynamic functions. Such quantities are invariant under $\sigma$ since it is absolutely continuous with respect to the Lebesgue measure. This result is somewhat surprising in the case of the nonergodic regime of the PM map (\ref{PMmap}), showing once more the strength of the ADK theorem.

From a dynamical point of view, the thermodynamic formalism allows us to obtain important quantities that characterize nonlinear systems. We can mention, for instance, the Pesin formula relating Lyapunov exponent $\Lambda$ to the Kolmogorov-Sinai entropy $h_{1}=h_{KS}$, namely $h_{KS}=[P(\beta)-P'(\beta)]_{\beta\rightarrow1_{-}}=\Lambda$ \cite{BS}. For nonergodic regimes $0<\alpha<1$, however, the topological pressure (\ref{PTph}) gives us the trivial relation $h_{KS}=\Lambda=0$. In fact, the dynamic instability is stretched exponential for such cases, of the form $|\delta x_{t}|\sim|\delta x_{0}| \exp(\lambda_{\alpha} t^{\alpha})$, rather than exponential $\alpha=1$. For such cases we can consider, once more, the infinite ergodic theory approach. It has recently been shown that the Pesin relation can be extended in a nontrivial way provided one introduces a convenient subexponential generalization of the Lyapunov exponent and Kolmogorov-Sinai entropy \cite{SV}. Moreover, the generalizations of such quantities behave like Mittag-Leffler random variables with $h_{\alpha}$ as the first moment \cite{SV,PSV}. A quest for new constitutive relations involving $P(\beta)$ that lead directly to these results in the thermodynamic formalism probably deserve further investigations.

\acknowledgments

The author thanks A. Saa for enlightening discussions. This work was supported by the Brazilian agency CNPq.


\begin{thebibliography}{99}
\bibitem{note} From the series definition of the Mittag-Leffler function, it is easy to see that $E_{\gamma}(-u)=2E_{2\gamma}(u^{2})-E_{\gamma}(u)$, thus $E_{\gamma}(u)\sim0$ for $u\rightarrow-\infty$.
\end{thebibliography}
\end{document}